\documentstyle[11pt,aaspp4]{article}
\newcommand\beq{\begin{equation}}
\newcommand\eeq{\end{equation}}

\begin{document}

\title{Emission Spectra of Fallback Disks Around Young Neutron Stars}
\author{Rosalba Perna\altaffilmark{1}, Lars Hernquist and Ramesh Narayan}
\medskip
\affil{Harvard-Smithsonian Center for Astrophysics, 60 Garden Street,
Cambridge, MA 02138}

\altaffiltext{1}{Harvard Society of Fellows}

\begin{abstract}

The nature of the energy source powering anomalous $X$-ray pulsars is
uncertain.  Proposed scenarios involve either an ultramagnetized neutron
star, or accretion onto a neutron star. We consider the accretion
model proposed recently by Chatterjee, Hernquist \& Narayan, in which
a disk is fed by fallback material following a supernova.
We compute the optical, infrared, and submillimeter emission expected
from such a disk, including both viscous dissipation and
reradiation of $X$-ray flux impinging on the disk from the pulsar. We
find that it is possible with current instruments to put serious
constraints on this and on other accretion models of AXPs.  Fallback
disks could also be found around isolated radio pulsars and we compute
the corresponding spectra. We show that the excess emission in the R
and I bands observed for the pulsar PSR 0656+14 is broadly consistent
with emission from a disk.

\end{abstract}


\section{Introduction}

In the past few years, the realm of astronomical objects has been
enriched with the discovery of about half a dozen of the so-called
anomalous $X$-ray pulsars (AXPs) (Mereghetti \& Stella 1995; van
Paradijs, Taam \& van den Heuvel 1995).  AXPs are sources of pulsed
$X$-ray emission, with persistent luminosities $L_x\sim 10^{35} -
10^{36}$ erg ${\rm s}^{-1}$ and soft spectra. Their periods lie in a very narrow
range, between 6 and 12 seconds and their characteristic ages are 
of order $10^3 - 10^5$ years.  To date, about half the known
AXPs have been associated with Galactic supernova remnants.  (For 
a recent review of the properties of AXPs, see Mereghetti
1999.)

While AXPs are almost certainly neutron stars, their properties
establish them as a distinct class from  binary $X$-ray pulsars
and radio pulsars.  Compared to binary $X$-ray pulsars, AXPs have
lower luminosities and exhibit a narrow distribution of periods.
Unlike young radio pulsars, AXPs have rather longer periods and appear to be
radio quiet.  One key to understanding these differences is to try
and identify the energy source which powers the $X$-ray
emission.  It is quite clear that this energy cannot be provided by
rotation (as it is in radio pulsars).
For values of $P$ and $\dot{P}$ that are characteristic of
AXPs, the rate of loss of rotational energy is $|\dot{E}|\equiv 4\pi^2
I \dot{P}/P^3 \approx 10^{32.5}$ erg ${\rm s}^{-1}$, orders of magnitude smaller
than the observed $X$-ray luminosities.

Two broad classes of models have emerged to account for the $X$-ray
emission from AXPs.  In one, AXPs are hypothesized to be isolated,
ultramagnetized neutron stars, with field strengths in the range
$10^{14} - 10^{15}$ G, a.k.a. ``magnetars'' (Duncan \& Thompson 1992).
Field strengths of this order are consistent with those inferred for
soft gamma repeaters, on the basis of timing measurements, assuming
that these objects lose rotational energy in a manner similar to radio
pulsars (e.g. Kouveliotou et al. 1998, 1999).  If the magnetar
interpretation of AXPs is correct, the $X$-ray luminosities could be powered
by either magnetic field decay (Thompson \& Duncan 1996) or by
residual thermal energy (Heyl \& Hernquist 1997a,b).  The observed
$X$-ray luminosities place constraints on both mechanisms.  If the
emission is powered by residual thermal energy, then the envelope of
the star must consist primarily of light elements such as hydrogen and
helium (Heyl \& Hernquist 1997b). On the other hand, if the emission
is powered by magnetic field decay, then a value of $B\ga 10^{16}$ G
is required, unless non standard decay processes are invoked (Heyl \&
Kulkarni 1998).

The second class of models proposes that the $X$-ray emission from
AXPs is powered by accretion, and unusually high values of the
magnetic field are not required.  Accretion can occur from binary
companions of very low mass (Mereghetti \& Stella 1995), from the
debris of a disrupted high mass companion (van Paradijs, Taam \& van
den Heuvel 1995; Gosh, Angelini \& White 1997) or from the
interstellar medium (Wang 1997).  Each of these three possibilities
has been challenged by observational and theoretical considerations.
Searches to detect binary companions of AXPs have failed thus far,
placing severe constraints on companion masses (e.g. Wilson et
al. 1998; Mereghetti, Israel \& Stella 1998).  If the emission is
powered by accretion from a disrupted binary companion, it is not
clear why AXPs should be associated with young supernova remnants.
Finally, for the specific object (RXJ 0720.4-3125) considered by Wang
(1997), Heyl \& Hernquist (1998) have demonstrated that the $X$-ray
luminosity can be explained by cooling, without requiring that this
neutron star be old and have experienced magnetic field decay.

Recently, Chatterjee, Hernquist \& Narayan (1999; hereafter, CHN) have
studied another accretion model, in which neutron stars accrete from
disks which formed from fallback material after a supernova explosion.
(Similar proposals have been studied recently by Alpar [1999] and
Marsden et al. [1999].)
The possibility that material might fall back onto a neutron star
following a supernova explosion and settle in a disk is not new
(e.g. Woosley 1988, Chevalier 1989).  In particular, Lin, Woosley \&
Bodenheimer (1991) suggested that this process might account for the
presence of planets around some radio pulsars.  Later, in their
discussion of accretion from the debris of a disrupted companion, van
Paradijs et al. (1995) noted that fallback could provide material to
form disks around young AXPs, but they did not analyze this
possibility in detail.

In principle, all neutron stars could acquire debris through fallback,
but subsequent accretion can occur only under specific circumstances,
depending on the relative locations of the magnetospheric radius, the
light cylinder radius, and the corotation radius.  CHN related these
conditions to physical parameters such as the initial mass of the
disk, $M_{\rm d}$, the initial period of the neutron star, $P_0$, and
the strength of the magnetic field, $B$.  CHN found that neutron stars
would evolve into accreting sources only for sufficiently strong
magnetic fields, while objects with weaker fields would be unaffected
and would, presumably, become ordinary radio pulsars.  For the
particular values $P_0 = 0.015$ seconds and $M_d = 0.006 M_\odot$, for
example, CHN found that the outcome depended on whether or not the
field is greater or smaller than $B = 4\times 10^{12}$ G. Namely,
neutron stars with $B> 4\times 10^{12}$ G became AXPs, while those with
$B< 4\times 10^{12}$ G became radio pulsars. Other choices for
$P_0$ and $M_{\rm d}$ yield similar behavior, and lead to different
values for the critical magnetic field separating AXPs and radio
pulsars.  

In their analysis, CHN demonstrated that the time varying nature of
accretion from a fallback disk provides a natural explanation for the
fact that AXP periods increase with time.  Indeed, if the
accretion rate declines as $\dot{M} \propto t^{-7/6}$, close to the
similarity solution derived by Cannizzo, Lee \& Goodman (1990), then
the rate of increase of the period will be identical to spindown by
magnetic dipole radiation.  Thus, timing analyses of steady spindown
alone cannot distinguish between these two interpretations.  Moreover,
in contrast with the magnetar model, the time varying accretion
studied by CHN relies on a standard population of isolated neutron
stars having magnetic fields similar to those inferred for radio
pulsars, and therefore it does not require the existence of a separate
class of neutron stars.

How can we distinguish between the time-dependent
accretion interpretation and the magnetar model, given that steady spin-down is
accounted for by both?  One possibility may be by searching directly for 
disk emission, at wavelengths much longer than $X$-rays. 
Accretion disks dissipate energy by
viscous processes at all radii, and some of this energy will be radiated
at long wavelengths, especially from large radii. 
Moreover, the central $X$-ray emitting source irradiates the outer disk, and
most of this energy is reradiated at longer wavelengths.  In this
paper, we compute the emission spectrum expected for accretion disks
like those required to explain AXPs, and we apply the model of CHN to
make predictions for the flux in various bands.
The predicted flux is within the range of sensitivity of current
instruments, and can be used, therefore, to directly test the
accretion model.  We note that while our specific estimates are made
using the disk models of CHN, our arguments should apply in general to
all accretion models for AXPs regardless of the source of the disk mass.
Thus, sensitive observations at long wavelengths may provide powerful
tests of all variants of the accretion hypothesis for AXPs.

A prediction of the CHN analysis is that fallback disks could also be
found around radio pulsars, although the properties of these disks are
not well-constrained.  Consequently, we apply our methods to calculate
long wavelength emission from disks around radio pulsars.  In
particular, we consider the middle-aged radio pulsar PSR0656+14, which
has emission in the R and I bands in excess of that expected by
extrapolating the thermal blackbody spectrum observed at $X$-ray
wavelengths. We show that the excess emission is broadly consistent
with that produced by an accretion disk.

\section{Model}

The evolution of the fallback material settling into a disk can be
treated in a similar manner to the evolution of the accretion disk
formed by material captured by a black hole after the tidal disruption
of a star.  Numerical simulations by Cannizzo, Lee \& Goodman (1990;
hereafter CLG) show that after an initial transient phase, the 
evolution of the disk becomes self-similar, with the accretion rate
$\dot{M}$ decreasing as a power law. Exact self-similar solutions for 
the evolution of the disk were derived by Pringle (1974) assuming that
$\nu\propto r^p \Sigma^q$, where $\nu$ is the viscosity and $\Sigma$ 
the surface density.  If the
so-called $\alpha$ disk model is adopted (see CLG for details), then a
solution is obtained when $p=1$ and $q=2/3$, yielding \beq
\nu=C\,r\,\Sigma^{2/3}\;,
\label{eq:nu}
\eeq
with
\beq
C\equiv\alpha^{4/3}\left(\frac{k_B}{\mu m_p}\right)^{4/3}
\left(\frac{\kappa_{\rm es}}{12acGM}\right)^{1/3}\;.
\label{eq:C}
\eeq
Here $\kappa_{\rm es}$ is the electron scattering opacity
\footnote{CLG considered more general opacity laws, such as the
Kramers opacity, with $\kappa(\rho ,T)=\kappa_0\rho T^{-7/2}$, but they
found their results to be relatively insensitive to the particular
expression they chose.  Therefore, for simplicity we adopt a
constant opacity, $\kappa_{\rm es}$.}, $M$ the mass of the neutron star, and $\alpha$
is the standard 
dimensionless viscosity constant that is expected to be $\la 1$ on physical
grounds (Shakura \& Sunyaev 1973). We adopt the value $\alpha=0.1$
in our numerical calculations.
The corresponding surface density evolves according to
\beq
\Sigma(r,t)=\Sigma_0\left(\frac{t}{t_0}\right)^{-15/26}
f\left[\left(\frac{r}{r_0}\right)\left(\frac{t}{t_0}\right)^{-3/8}\right]\;,
\label{eq:sigma}
\eeq
where
\beq
f(u)\equiv(28)^{-3/2}u^{-3/5}(1-u^{7/5})^{3/2}\;.
\eeq
The constants $\Sigma_0$, $r_0$ and $t_0$ are constrained by the condition
\beq
t_0^{-1}=\frac{C\Sigma_0^{2/3}}{r_0}\;.
\label{eq:const}
\eeq
The total mass of the disk decreases with time as
\beq
M_{\rm d}(t)=(28)^{-3/2}\frac{4\pi}{7}r_0^2\Sigma_0
\left(\frac{t}{t_0}\right)^{-3/16}\;.
\label{eq:md}
\eeq 
The accretion rate is then given by $\dot{M_{\rm d}}(t)=(-3/16)\,M_{\rm d}(t)/t$.
We define the initial mass of the disk to be $M_{\rm d}^0\equiv
M_{\rm d}(T_d)$, where $T_d$ is of order the dynamical time of the
disk early on. Following CHN, we select $T_d=1$ msec.
Having chosen a value for $r_0$, the constant $\Sigma_0$ is then derived from
equation~(\ref{eq:md}) at $t=T_d$. Using the relation (\ref{eq:const}),
we then find 
\beq 
{\Sigma_0}\approx 155\;(M_{\rm d}^0)^{8/7}\;(T_dC)^{3/14}\;r_0^{-5/2}\;.  
\label{eq:sig0}
\eeq 
Notice that, with the choice (\ref{eq:sig0}), (i.e. $\Sigma_0\propto
r_0^{-5/2}$), the self-similar solution for $\Sigma$, as given by
Eq. (\ref{eq:sigma}), becomes independent of the numerical value of the
parameter $r_0$, and therefore is solely determined by $M_{\rm d}^0$
and $T_d$. However, as Equation (\ref{eq:sig0}) shows, the dependence
on $T_d$ is much weaker than the dependence on $M_{\rm d}^0$
\footnote{Large variations in the choice of $T_d$ would lead to the
same numerical results with small adjustments in $M_{\rm d}^0$.}. Following
CHN, we keep the value of $T_d$ fixed and use only $M_{\rm d}^0$ as a parameter. 

The solution (\ref{eq:sigma}) extends from $u=0$ to $u=1$. However, an
accretion disk can exist only for radii beyond the magnetospheric
radius $R_m\approx 6.6\times 10^7 B_{12}^{4/7}\dot{m}^{-2/7}$ (Frank,
King \& Raine 1992), where $B_{12}\equiv B/(10^{12}G)$ and
$\dot{m}\equiv \dot{M}/\dot{M_E}$ is the mass accretion rate in
Eddington units. We therefore take the minimum radius of the disk to be $r_{\rm
min}=R_m$. Similarly, whereas the mathematical solution sets the disk
outer boundary at $u=1$ (corresponding to $\Sigma = 0$), in practice a
disk can extend only up to a radius $r_{\rm max}(\Sigma_{\rm min})$,
where densities $\Sigma \la \Sigma_{\rm min}$ will be destroyed by the
ram pressure from the interstellar medium.  We assume that 
a disk can exist as long as the incident flux is not
sufficient to significantly perturb the angular momentum of the matter
orbiting in it. We find that this condition is satisfied for
densities $\Sigma \ga \Sigma_{\rm min}\approx
5\times 10^{-7} r_{14}^2\; M_{1.4}^{-1} \;n_0\; v^2_{300} \;{\rm
g}\;{\rm cm}^{-2}$ where $r_{14} \equiv r/(10^{14} \, {\rm cm})$,
$M_{1.4} \equiv M/1.4 M_\odot$, $n_0 \equiv n/$ (1 gm/cm$^{3}$) is the
particle density in the interstellar medium, and $v_{300} \equiv v /
300$ km/sec.  Therefore interstellar ram pressure affects the outer
boundary of the disk only very marginally.

At any position and time, the effective temperature of the disk due    
to viscous dissipation is given by (CLG)
\beq
T_{\rm eff}(r,t)=\left(\frac{9\nu(r,t)\Sigma(r,t)GM}
{8\sigma r^3}\right)^{1/4}\approx \left(-
\frac{3\dot{M_{\rm d}}(t)GM}{8\pi\sigma r^3}\right)^{1/4}\;.
\label{eq:t1}
\eeq
Besides emission due to internal energy dissipation, a considerable
amount of flux from the disk can arise as a result of reradiation of 
$X$-rays impinging from the central object. Vrtilek et al. (1990) derive an
analytical expression for the effective temperature of an
irradiated disk, under the assumption that the disk height
$h\propto r^n$ and irradiation is the dominant form of heating.
 If $L_X$ is the $X$-ray luminosity of the central source, 
and $\omega\equiv \sqrt{GM/r^3}$ is the
Keplerian rotation frequency of the disk at position $r$,
then $n=9/7$, and the temperature profile is given by
\beq
T_X(r)=\left[f\frac{\sqrt{k_B/\mu m_{\rm H}} L_X \omega}
{14\pi\sigma GM}\right]^{2/7}  
\simeq 23200
\left(\frac{f}{0.5}\right)^{2/7}\left(\frac{L_X}{L_E}\right)^{2/7}
\left(\frac{R_\odot}{r}\right)^{3/7}\;.
\label{eq:t2}
\eeq
The last equation is for $M=1.4M_\odot$. Here, $f$ is a factor which
contains the uncertainty in the disk structure (including the absorbed
fraction of radiation impinging on the surface). We adopt the estimate
given by Vrtilek et al. (1990) of $f\sim0.5$.  In the cases that we
consider, irradiation typically dominates the spectrum for $L_X\ga
10^{34}$ erg s$^{-1}$; for lower values of $L_X$ (but generally higher
than $\sim 10^{31}$ erg s$^{-1}$), 
the inner parts of the disk (where the optical emission is mostly
generated) are dominated by viscous heating, while irradiation 
dominates in the outer regions of the disk (where most of the
longer-wavelength emission comes from). Therefore it is reasonable to
keep the approximation (\ref{eq:t2}) also in these cases and compute
the contribution to the flux of the two terms separately.

The total flux is obtained by integrating the emissivity over the
entire surface of the disk:
\beq
F=\frac{1}{d^2}\int_{r{\rm min}}^{r{\rm max}}dr\,r \int_0^{2\pi}d\phi
[B_\nu(T_{\rm eff}) + B_\nu(T_{\rm X})]\times
[\cos\phi \sin\beta(r) \sin\alpha + \cos\beta(r) \cos\alpha]\;.
\label{eq:flux}
\eeq
Here $\alpha$ is the angle between the line of sight and the normal 
to the disk at $r=0$, and $\beta(r)=dh(r)/dr$ is the tilt angle of the
surface of the disk, which we determine at every position in the disk from the
equation of hydrostatic equilibrium in the z-direction.

The opacity in disks made of fallback material is
likely to be higher than in disks with cosmic abundance, due
to the expected high metal content. Following Lin, Woosley \& Bodenheimer
(1991), we adopt an opacity of $\sim 10^3 {\rm cm}^2 {\rm g}^{-1}$ in
our calculations. Notice, however, that in our model the predicted spectrum is
insensitive to the precise value of the opacity, as 
the temperature is independent of it. 
The size of the disk is slightly larger for high $\kappa_{\rm es}$,
but this has little influence on the final result. 
What is important, however, is that a high opacity ensures that
the disk is optically thick at all radii, and therefore
the emission can be computed as blackbody.
A more significant uncertainty derives from the possible existence
of thermally unstable regions in the disk. For disks with cosmic
abundance, the unstable temperature ranges have been studied in
detail (i.e. Cannizzo 1993 and references therein). For high
metallicity disks the situation is likely to be quite different, as 
cooling depends on the abundances. 

\section{Emission from disks around AXPs and radio pulsars}

In the model that we consider (CHN), AXPs and radio pulsars are
assumed to be drawn from the same underlying population of neutron
stars. 
In one example considered by CHN, for a fallback
disk with mass $M_{\rm d}^0=0.006M_\odot$, and an initial period
$P_0=15$ msec, the evolution of the accreting disk will lead to an AXP
phase if the magnetic field of the neutron star is $B_{12}\ga
4$. Smaller values of the magnetic field would lead to the more common
radio pulsar phase.  In the following, we consider these two
situations separately, and compute the emission spectrum from the disk
at various ages.

\subsection{AXPs}

Here we consider a disk of mass $M_{\rm d}^0=0.005M_\odot$
at an angle of $60^o$ with respect to the line of sight, and a
magnetic field $B_{12}=8$ for the neutron star. 
According to CHN, such objects will go through 
a phase of accretion-induced $X$-ray emission and will appear as AXPs. 
We describe the $X$-ray luminosity as 
\beq
  L_X = \left\{
  \begin{array}{ll}
   \dot{m}\, L_{\rm E}  & \hbox{if $\dot{m}\ge 0.01$} \\
    100 \,\dot{m}^2  & \hbox{if $\dot{m}\le 0.01$}   \\
  \end{array}\right.\;.
\label{eq:lx1}
\eeq
The rapid decline of the luminosity for $\dot{m}< 0.01$ is to allow for the fact
that the system is likely to switch to an ADAF phase, at which point the
accretion luminosity $L_X$ decreases faster than $\dot{M}$
(Narayan \& Yi 1995; Esin et al 1997). Some
studies have suggested, in fact, that in this phase much of the mass could be
ejected prior to reaching the surface of the star (e.g. Blandford
\& Begelman 1999, Menou et al. 1999, Quataert \& Narayan 1999). 
The transition radius between the ADAF and the disk is not well
determined. Moreover, it is likely to depend on the metallicity of the
gas in the disk, which is very far from the standard value. For simplicity,
we keep the transition radius equal to the Alfven radius (which at later
times is on the order of a few thousand Schwarzschild radii).
We will discuss the consequences of this uncertainty on our results.

Figure 1 shows the unextincted flux at different ages from an AXP at a
distance of 5 kpc.  The peak in the submillimeter range is due to the
effect of irradiation. At early times ($T\la 10^4$ yr), irradiation dominates
over viscous dissipation in 
the entire disk.  At later times, when the $X$-ray flux is
much lower, viscous heating becomes dominant in the inner parts of the
disk, and this produces an enhancement in the emission 
in the optical region of the spectrum. The model predicts a flux
of a few mJy at frequencies around $10^{12}$ Hz, which is within the range of
the SCUBA instrument.  The infrared emission is also generally above
the sensitivity limit of NICMOS on the Hubble Space Telescope. 
In the J band (1.1 $\mu$m), the limiting
magnitude for a 5$\sigma$ detection in 60 minutes of exposure is
mag $\sim 25$, which corresponds to a flux of $\sim 10^{-4}$ mJy =
$10^{-30}$ erg cm$^{-2}$ s$^{-1}$ Hz$^{-1}$. 

Presently, about half a dozen AXPs are known.
In Table 1 we summarize the properties of the nearest ones,
and compute the expected emission in various bands, assuming that the
accretion model is correct. As an illustration, we take $M_{\rm d}^0=0.005M_\odot$,
$B_{12}=8$, and an inclination angle of the disk $i=60^0$ in all cases. 
 $X$-ray luminosity, age and distance are taken
from observations\footnote{The observed $X$-ray luminosity is consistent
with the value predicted by the accretion model of CHN.}, 
and extinction by dust has been taken into
account based on the measured column density of hydrogen and
an average extinction for the Galaxy\footnote{Needless to say, 
given the uncertainties in both theoretical and
measured parameters, the numbers in the table must be simply
considered as rough estimates}. 
Following Pei (1992), we fit the extinction curve by 
\beq
A(\lambda) = k\xi(\lambda)\left(\frac{N_{\rm H}}{10^{21}{\rm cm}^{-2}}
\right)\;,
\label{eq:ext}
\eeq
where
\beq
\xi(\lambda) = \sum_{i=1}^6\frac{a_i}{(\lambda/\lambda_i)^{n_i} +
(\lambda_i/\lambda)^{n_i} + b_i}\;.
\eeq
Here $a_i,b_i,n_i,\lambda_i$ are fitting parameters (that can be found
in Pei 1992), and $k$ is the dust-to-gas ratio. For the Galaxy, $k=0.78$.

Observations in some bands have already been made for a couple of the
objects in the table.  In the case of 1E 2259, our predicted flux is
below the upper limits set by Coe \& Pighling (1998) in the J band
($\ga 19.6$), and is somewhat higher than the limit of $\sim$ 18 mag
set in the K band. For the case of 4U 01412+615 reported by Mereghetti
(1999), we get a flux comparable to the flux limit of 25 mag in the V
band, and somewhat higher than the limit of 17 mag in the K band.  We
need to point out, however, that in our model the flux in the higher
energy part of the spectrum is rather sensitive to the choice of
$r_{\rm min}$, which we take equal to the Alfven radius. If the inner
bound of the disk is at a radius larger than this, then the optical
and near infrared emission will be lower. The submillimeter flux, on
the other hand, is produced at large radii, and therefore is quite
insensitive to the specific value of $r_{\rm min}$. All fluxes are
sensitive to $M_{\rm d}^0$, for which we have assumed the value 0.005
$M_\odot$ suggested by CHN. It is unclear how much freedom is
available in this parameter, but it is probably not very
large. Another uncertainty is the inclination angle of the
disk, for which we have assumed the value $i=60^0$.
Furthermore, the distances to most of these objects are
not well constrained. Given all these uncertainties, more
and deeper observations in several bands are needed before
drawing definitive conclusions.
 
Observations in the optical (R and B band) have been made for an
object not in the table, RXJ 0720.4-3125, which shares some
characteristics with the known AXPs (Kulkarni \& van Kerkwijk 1998),
such as a long period (8.39 sec), but which has a rather low
luminosity ($4\times 10^{32}$ ergs s$^{-1}$); there is no agreement on
whether it belongs to the class of AXPs. There is a claimed detection
of this source in R and B, but it is unclear whether the observed
emission is from RXJ 0720 or if it is instead due to a background
star. If it is the latter, then, for the estimated distance of 400 pc,
the accretion model of CHN requires the age of this object to be $\ga
2\times 10^6$ yr for the emission to be below the detected values.
Since no $\dot{P}$ has been measured, there is no other constraint on
the age. If we assume that the optical emission does come from RXJ
0720, then the CHN disk model would require an age around $2\times
10^6$ yr to produce an emission in roughly the same range. The slope
of the spectrum, however, is not consistent with the disk model; the
model predicts that the R band should have higher flux than the B
band, while the observations show the contrary.  Observations in other
bands are needed to confirm the spectral slope\footnote{Note,
incidentally, that also in the case of PRS 0656+14 (Fig. 3), the flux
in the B band does not follow the trend of the other data points.}.

\subsection{Radio pulsars}
In the accretion picture, 
AXPs and pulsars are drawn from the same
population, and thus it is reasonable to expect that disks from
fallback material should also be found around some pulsars.  Figure 2
shows the emission spectrum from a disk surrounding a pulsar at
various ages. The $X$-ray luminosity of the star at each age has been
taken from the cooling curve for a neutron star with an iron envelope
and $B=0$ computed by Heyl \& Hernquist (1998)
\footnote{In the model made by  Heyl \& Hernquist, various
cooling curves, corresponding to different 
choices of the composition of the envelope of the neutron star
and strength of its magnetic field, cross around $10^5$ yr. 
Therefore, except for either very young or very old systems, our
results are not very sensitive to this particular choice.}. In all cases, we
chose $B_{12}=3$, and a disk mass equal to $M_{\rm d}=0.005M_\odot$ at
an inclination angle of $60^o$, as for AXPs; however, 
whereas there is a relatively 
tight constraint on the specific value of the disk mass for AXPs, the disk
mass could be much lower (or indeed even zero) for radio pulsars. Thus,
our predictions are less firm. 

We have applied our model to PSR 0656+14. This is an isolated,
middle-aged pulsar which shows an enhanced emission redward of the V
band (Pavlov, Welty \& Cordova 1997). 
The data are shown in Figure 3. The soft X-ray part of the
spectrum is well fitted by a thermal component, originating from the
surface of the neutron star. This part of the spectrum looks similar
to that observed in other pulsars, such as Geminga. 
As Pavlov et al. notice, however, the excess
emission in the R and I bands is not very common, 
and they fitted it with a non-thermal component
assumed to be produced in the magnetosphere. Here we consider the
possibility that, whereas the high energy emission comes from the
surface of the star (and is well fitted by a Rayleigh-Jeans spectrum), 
the low-energy emission could be produced    
by a residual disk. Following the notation of Pavlov et al. for the
Rayleigh-Jeans part, we fit the spectrum as the sum of these two
components: 
\beq f(\nu) = \left[F_{\rm disk}(\nu) +
g_0\left(\frac{\nu}{\nu_0} \right)^2\right] \times 10^{-0.4A(\nu)}\;,
\label{eq:fit}
\eeq 
where $\nu_0$ is an arbitrary reference frequency (which we took
equal to $5.8\times 10^{14}$ Hz), $g_0=3.12\times 10^{-31}G$ ergs
cm$^{-2}$ s$^{-1}$ Hz$^{-1}$, and $G\equiv
T_6(R_{10}/d_{500})^2$. Given the age $\tau=1.1\times 10^{5}$ yr
(Pavlov, Stringfellow \& Cordova 1996), a distance of 760 pc (Taylor,
Manchester \& Lyne 1993), an extinction of $A_{\rm V}=0.15$ and
$E(B-V)=0.05$ (Pavlov, Stringfellow \& Cordova 1996), a magnetic
field $B_{12}=4.7$, and an $X$-ray
luminosity of $3.5\times 10^{32} (D/760\; {\rm pc})^2 (F/5\times
10^{-12}{\rm erg}{\rm s}^{-1}{\rm cm}^{-2}{\rm s}^{-1})$ erg s$^{-1}$
(Cordova et al. 1989, Finley, \"{O}gelman \& Kizilo\u{g}lu 1992), the
only free parameters of the disk model are the disk mass and the
inclination angle of the disk. Figure 3 shows the resulting spectrum
corresponding to three choices of these parameters. Notice, however, 
that the observations imply rather low values of
$\dot{M}$, and the best fit is obtained for
an extreme value of the inclination angle.

If the assumption of an accretion disk surrounding this pulsar is
correct, then it should be possible to observe it directly. Figure 4 shows
the predicted spectrum over a wide range of wavelength.
The emission in the submillimeter is within
the range of sensitivity of SCUBA.

\section{Conclusions}

We have computed the emission from time-dependent accretion disks
around neutron stars, established by fallback following a supernova
explosion. Such disks have been proposed as candidates for the energy
source of AXPs. We applied our model to predict the emission in the
optical, infrared and submillimeter bands for the known AXPs. The
predicted fluxes are within the range of sensitivity of current
instruments. We need to emphasize, however, that whereas our specific
estimates have been made using the disk models of CHN, the arguments should
apply in general to any accretion model for AXPs regardless of the source
of the disk mass.  Therefore, deep observations at long wavelengths
can be used to test all variants of the accretion hypothesis
for AXPs.

If accretion disks are indeed responsible for powering the $X$-ray
emission from AXPs, then they should also be found around typical
radio pulsars. We computed the emission for parameters typical of
radio pulsars and showed that emission from a disk might
explain the excess flux redwards of the B band observed in the pulsar
PSR 0656+014. Sensitive observations at longer wavelengths would
severely test this interpretation.

In addition to perhaps exploring the energy source powering AXPs,
accretion disks around young compact objects may have other observable
consequences. Planets have been found around at least one pulsar, PSR B1257+12
(Wolszczane \& Frail 1992; Wolszczan 1994). Fallback debris would appear
to be an ideal source of planet-forming material since it would naturally 
be metal-rich if it is drawn from the inner regions of the progenitor star
(e.g. Lin et al. 1991). In some cases, fallback may be sufficiently intense 
to yield a black hole remnant rather than a neutron star. Recently, a class
of radio-quiet $X$-ray point sources has been identified, consisting of
approximately ten members, the majority of which are associated with young
supernova remnants (e.g. Brazier \& Johnston 1999). Time variability
in the emission from the source in SNR RCW 103, for example, has motivated
the possibility that this source may be an accreting black hole rather
than a cooling neutron star (e.g. Gotthelf, Petre \& Vasisht 1999). Qualitatively,
accreting disks established by fallback around black holes should be similar
to those examined here. Thus, deep observations at optical, infrared and 
submillimeter wavelengths may also reveal the true nature of radio quiet point 
sources in supernova remnants. 

This work was supported in part by NSF grant AST 980686.

\newpage

\hspace{-0.5in}\begin{tabular}{|c|c|c|c|c|c|} \hline
object &  1E 2259+586 & 1RXS J170849-400910 & 4U 01412+615 & 1E 1048-5937
& 1E 1841-045  \\ & (1) & (2) & (3) & (4) & (5) \\ \hline
$P/2\dot{P}$ (yr) & 1.75$\times 10^5$ & 8500  & 6$\times 10^4$  & 5000  &
4000  \\ \hline
D (kpc) & 4 & 10 & 4 & 10 & 7\\ \hline
$N_{\rm HI} ({\rm cm^{-2}})$ & $9\times 10^{21}$ &$1.4\times 10^{22}$ &$10^{22}$
& $10^{22}$ &$2\times 10^{22}$  \\ \hline
$L_X$(erg/sec) & $8\times 10^{34}$ & $1.2\times 10^{36}$ & $10^{36}$ &
$5\times 10^{35}$ & $3.5\times 10^{35}$ \\ \hline \hline
$F_{\nu_1}$ (mJy) &30 &5 &90 &3 &4 \\ \hline
$F_{\nu_2}$ (mJy) &$0.1$ &$0.4$ &$1$ &$0.1$ & $0.1$ \\ 
K(mag)& $\sim 17$ &$\sim 16$& $\sim 15$& $\sim 17$& $\sim 17$\\ \hline 
$F_{\nu_3}$ (mJy) &$10^{-4}$ &$6\times 10^{-5}$ &$2\times
10^{-3}$& $2\times 10^{-4}$ &$2\times 10^{-6}$  \\ 
V(mag) &$\sim26$ &$\sim 27$ &$\sim 24$ & $\sim 26$ &$\sim28$  \\  \hline
\end{tabular}

\bigskip

{Table 1. --- Predicted flux at frequencies $\nu_1=2000$ GHz,
$\nu_2=1.36\times 10^{14}$ Hz (K band), $\nu_3=5.4 \times 10^{14}$ Hz
(V band) for some of the known AXPs. The model parameters adopted here
are the same as in Figure 1. References for these objects can
be found in Corbet et al. (1), Sugizaki et al. (2), Wilson et al.
(3), Osterbroek et al. (4), Gotthelf \& Vasisht (5).}

\begin{figure}[t]
\centerline{\epsfysize=5.7in\epsffile{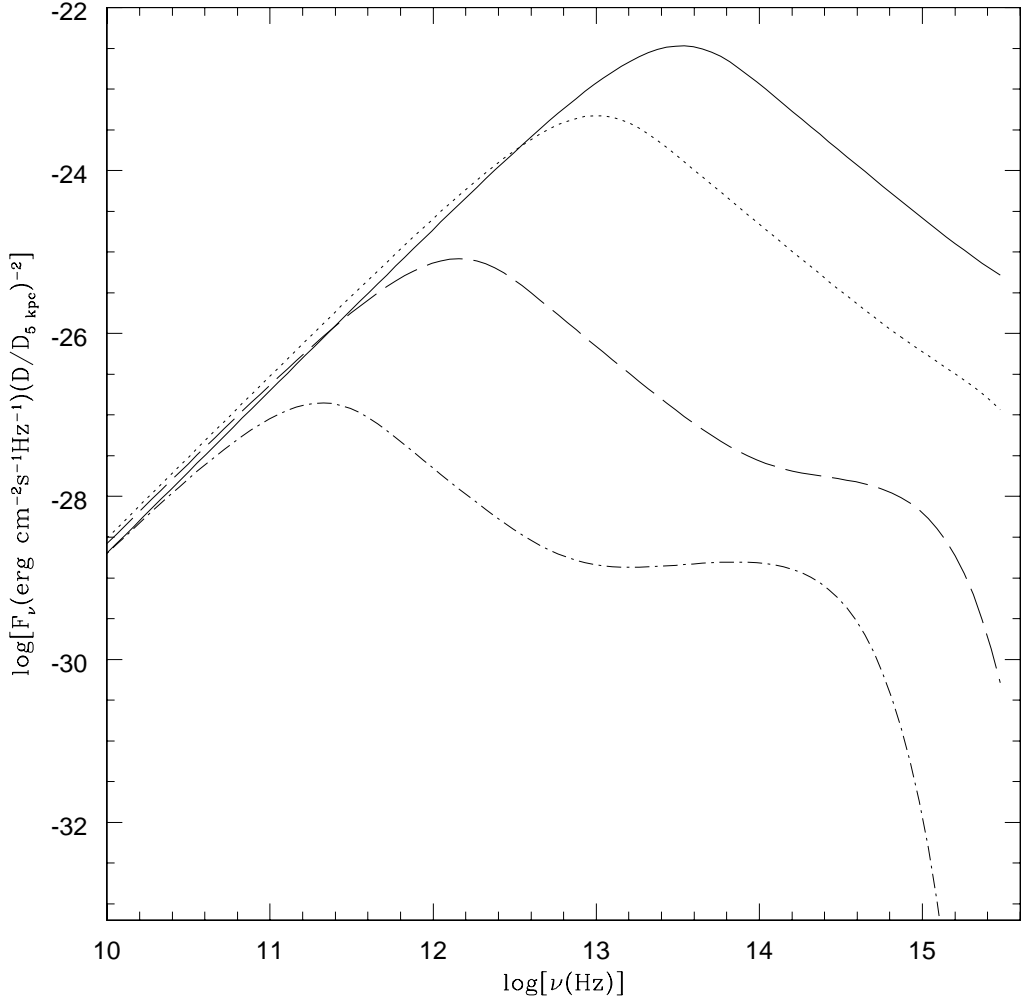}}
\caption{Emission spectrum from an accretion disk around an AXP at 
different ages. The disk mass is $M_{\rm d}^0=0.005 M_\odot$  
and the magnetic field strength is $B=8\times 10^{12}$ G, similar to
values considered in the accretion model of CHN. The irradiating $X$-ray
flux is assumed to derive from accretion onto the neutron star.
The four curves correspond to ages of $\tau=10^3$ yr
(solid line), $\tau=10^4$ yr (dotted line), $\tau=10^5$ yr
(dashed line), $\tau=10^6$ yr (dotted-dashed line). The inclination angle
of the disk is assumed to be $i=60^0$.}
\label{fig:1} 
\end{figure}

\begin{figure}[t]
\centerline{\epsfysize=5.7in\epsffile{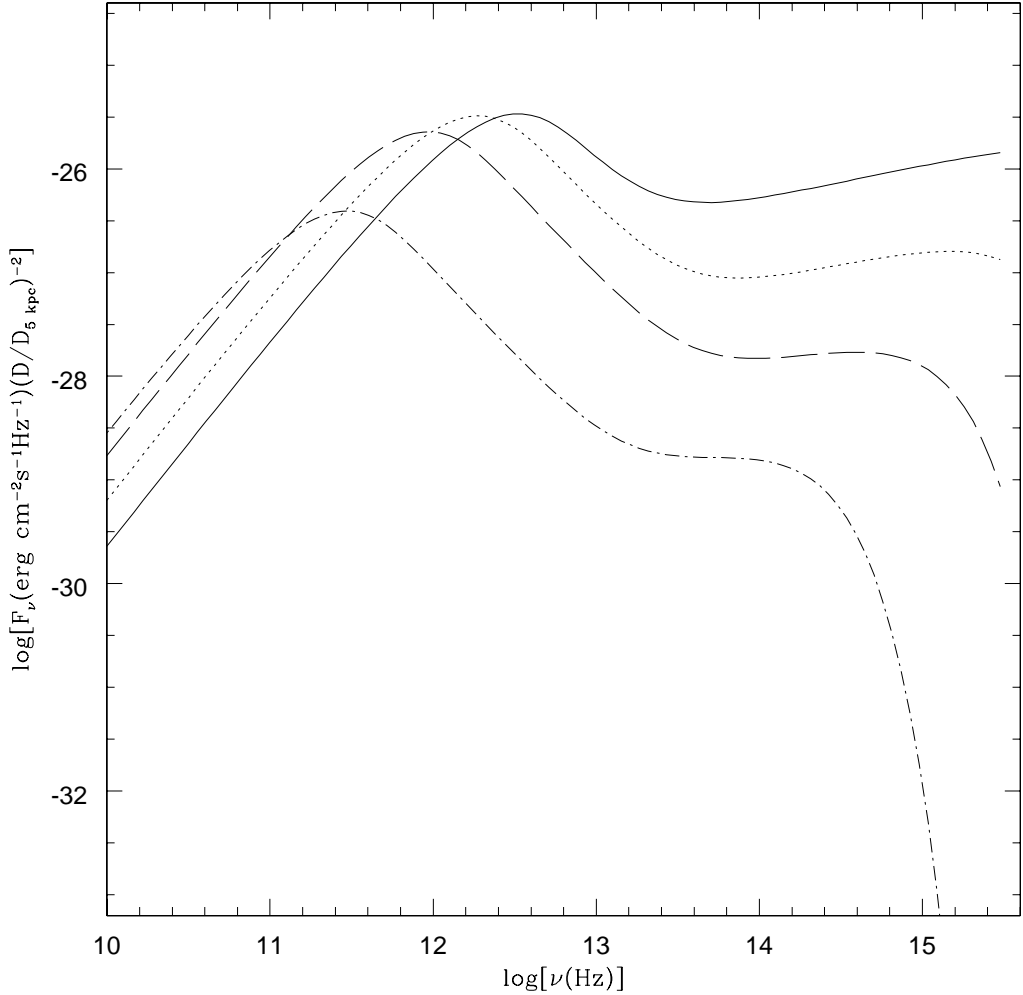}}
\caption{Emission spectrum from an accretion disk around a normal pulsar at 
different ages. Here the disk mass has the same value as in Fig. 1 for AXPs,  
but the magnetic field strength is smaller, $B=3\times 10^{12}$ G,
making accretion impossible according to the model of CHN. The irradiating $X$-ray
flux is assumed to be generated by cooling of the neutron star.
The four curves correspond to ages of  $\tau=10^3$ yr
(solid line), $\tau=10^4$ yr (dotted line), $\tau=10^5$ yr
(dashed line), $\tau=10^6$ yr (dotted-dashed line). The inclination angle
of the disk is assumed to be $i=60^0$.}
\label{fig:2}
\end{figure}

\begin{figure}[t]
\centerline{\epsfysize=5.7in\epsffile{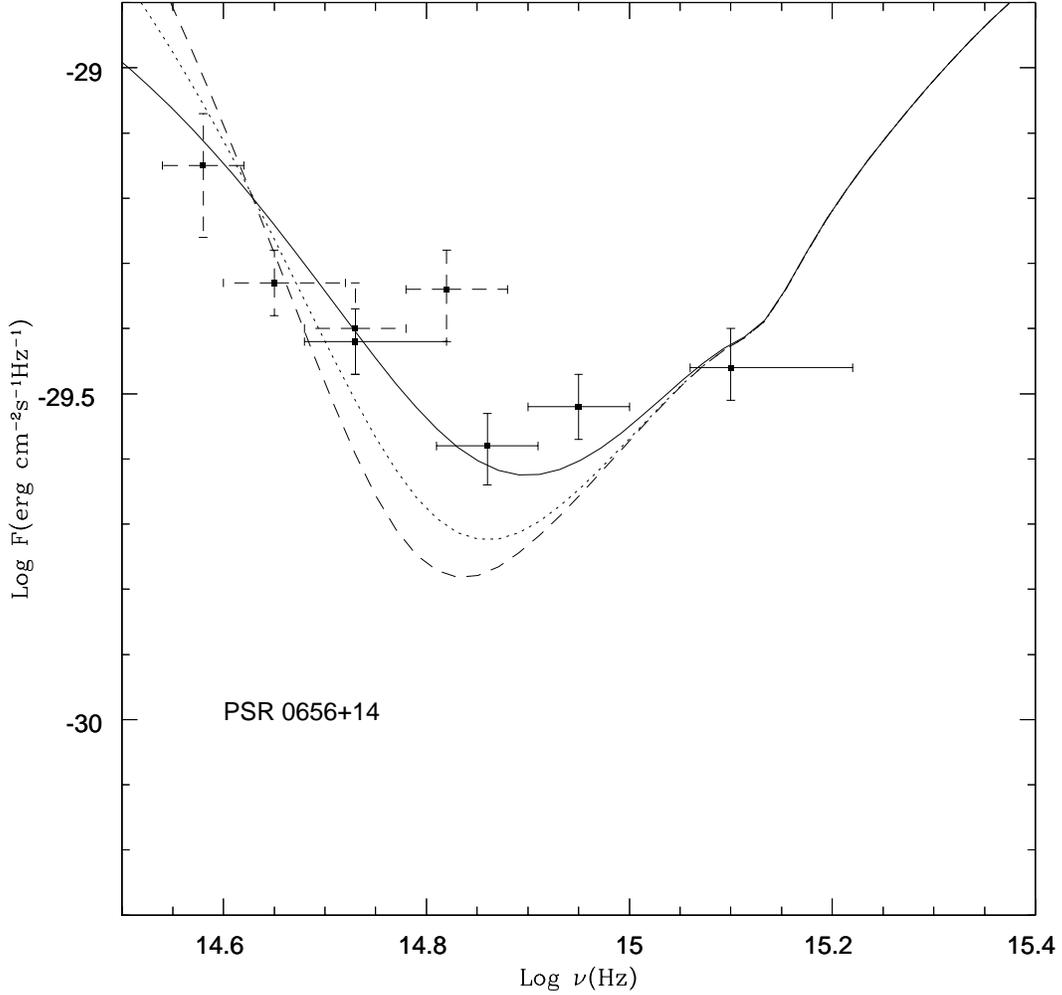}}
\caption{The excess flux redward of the B band observed for the pulsar
PSR 0656+14 is consistent with emission from an accretion disk.
Age, irradiating $X$-ray flux, dust extinction and distance are taken
from observations.  The various lines correspond to a model with
$M_{\rm d}^0= 0.00015 M_\odot$, $i=60^0$ (dashed line), $M_{\rm d}^0=
0.00023 M_\odot$, $i=80^0$ (dotted line), $M_{\rm d}^0= 0.0004 M_\odot$,
$i=87^0$ (solid line).  The emission at high energies is assumed to 
come from the surface of the neutron star.}
\label{fig:3}
\end{figure}

\begin{figure}[t]
\centerline{\epsfysize=5.7in\epsffile{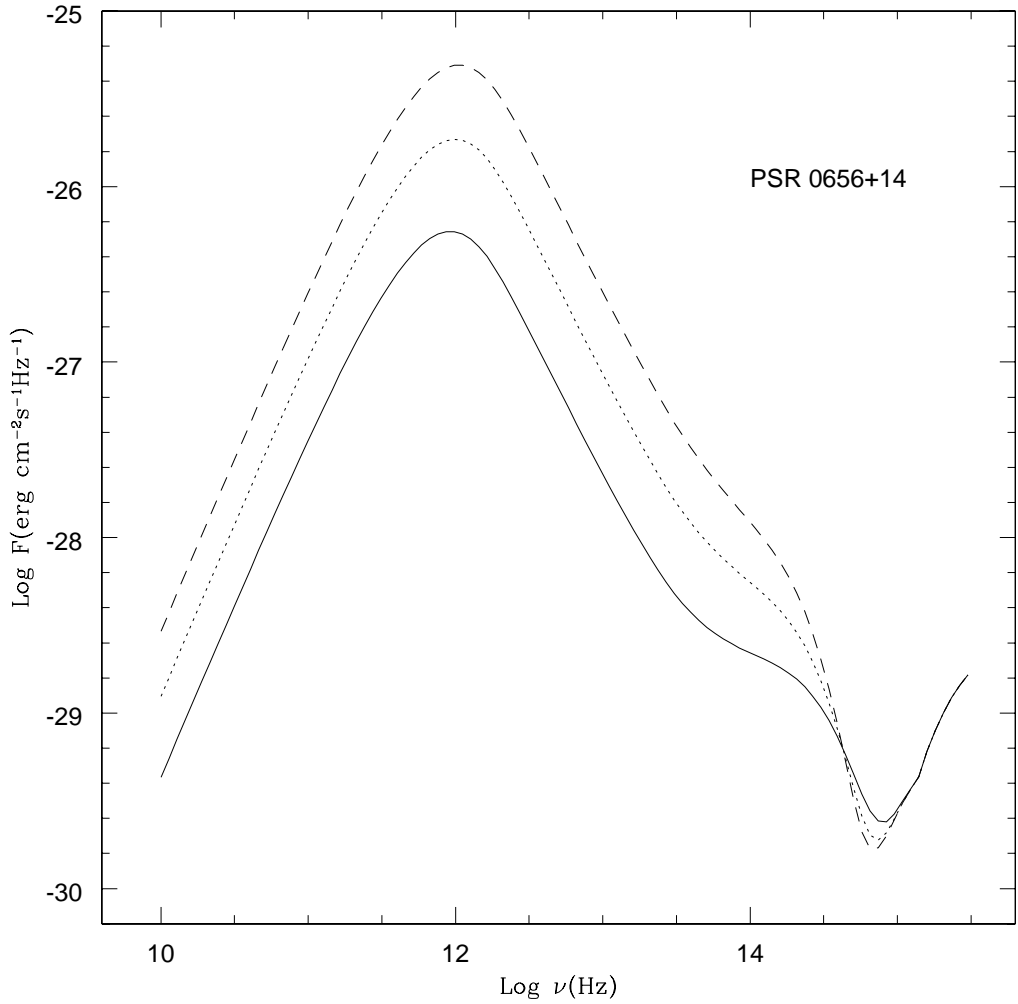}}
\caption{Emission spectrum from PSR 0656+14 if an accretion disk
is responsible for the excess emission observed redward of the B band.
The various lines correspond to the same models as in Figure 3.}
\label{fig:4}
\end{figure}

\end{document}